# Isotope effect in the work function of lithium

*Atef A. Sheekhoon*,[*] *Abdelrahman O. Haridy*,[*] *Vitaly V. Kresin*

Department of Physics and Astronomy, University of Southern California,
Los Angeles, 90089-0484, USA

May 2026

The work functions of $^{7}$Li and $^{6}$Li metals have been measured as a function of temperature, by using photoionization of pure isolated metal nanoparticles in a beam. These data reveal a marked isotope effect in the temperature variation of these work functions. Furthermore, for both isotopes the curvature of this temperature variation is found to be significantly larger than may be ascribed purely to a change in the electron gas density. These findings enhance the characterization of lithium as a quantum material in which the interplay between electronic and ionic degrees of freedom is nontrivial, and call for a microscopic understanding beyond simple models. Additionally, the slope of the work function curves was observed to vanish in the low temperature limit, as had been predicted on the basis of the Third Law of thermodynamics.

*These authors contributed equally to this work.

## I. INTRODUCTION

The interaction between metal conduction electrons and thermal lattice vibrations leads to well-known consequences for electronic kinetic phenomena, such as charge and heat transfer. But interestingly, the presence of vibrational excitations also can affect equilibrium properties of electron spectra. Specifically, here we probe the influence of thermal vibrations on the electronic work function. While the temperature dependence of the work function is not as steep as that of transport coefficients, an accurate measurement of this effect elucidates a distinctive connection between electronic and vibrational degrees of freedom. To spotlight the role of the latter, we focus on the isotope effect by comparing the temperature variations of the work functions of $^6$Li and $^7$Li.

In addition to having two abundant stable isotopes with sufficiently separated masses, lithium metal possesses other features that make it an interesting subject. The low atomic mass, substantial zero-point motion, and high Debye temperature (320–440 K [1,2]) result in a significant enhancement of quantum effects in its lattice structure and dynamics, giving rise to a range of unusual structural and phase transitions and even superconductivity with an anomalous isotope effect [3,4].

Alongside the low mass, lithium is distinguished by the relatively weak shielding of the atomic nucleus by the 1s core electrons, exposing its conduction electrons to a hard and nonlocal pseudopotential [5-7]. As a result, for example, their optical and thermal effective masses are much more enhanced relative to the free-electron value than in the other alkali metals [8]. Now consider that the work function can be expressed as follows [9,10]:

$$W = -e\Delta\varphi - \bar{\mu} \qquad (1)$$

where $e$ is the magnitude of the electron charge, $\bar{\mu}$ is the electron chemical potential inside the metal referred to the electrostatic potential in the interior, and $\Delta\varphi$ is the difference between the latter and the electrostatic potential in the vacuum outside the metal. This implies that for lithium one may anticipate that quantum effects will influence the dependence of $W$ on the temperature and on the isotopic mass: as the material warms up and undergoes thermal expansion, not only does the electron gas density change but so do the band structure, the electron-vibrational coupling strength, the effective mass, the behavior of the surface dipole layer, etc. As will be shown below, simple electrostatic models that can reproduce $W(T)$ for other metals aren't as successful for Li, which underscores the need for a more quantitative theory.

While lithium is a very interesting material to study, many of its physical and chemical properties can be impacted by contamination. Despite sometimes being characterized as "the least reactive" of the alkali group [11], it is the only alkali metal that significantly reacts with nitrogen, and in the molten state vigorously attacks glass, ceramics, copper, and graphite. Contaminants significantly influence spectroscopic and structural measurements [12,13] and can strongly distort work function values even in minute presence on the sample surface [14,15]. Since $dW/dT \sim 10^{-4}$ eV/K, precise and contamination-free measurements are necessary to trace out the $W(T)$ curve and resolve the isotopic variations.

We accomplish this by determining the work functions via photoionization of free isolated nanoparticles [16]. Their short flight time within the nanocluster beam apparatus and small surface area ensure the absence of contamination; then a careful measurement of their near-threshold photoionization provides the requisite fraction-of-a-percent precision in assigning the ionization energy over a wide temperature range.

## II. METHODS

The apparatus, described in detail in [16], produces metal nanoparticles using a gas aggregation source. A resistively heated stainless-steel crucible is loaded with oxide-free lithium metal, either naturally abundant $^7$Li (6.94 standard atomic weight, Sigma-Aldrich) or 95% purity $^6$Li (6.02 standard atomic weight, Sigma-Aldrich ISOTEC). The resulting metal vapor is quenched within a continuous flow of ultrahigh purity helium that is cooled by a liquid nitrogen-filled jacket surrounding the aggregation region. Transported by the helium gas, the metal vapor condenses into pure nanoparticles which are then carried through a "thermalization tube" where they are equilibrated to an internal temperature adjustable between 60 K and 360 K, with an accuracy of approximately 1 K [16,17]. Finally, the nanoparticles exit into vacuum as a directed beam.

This beam is populated by a distribution of nanoparticle sizes, which is characterized by time-of-flight mass spectrometry (Appendix A). The distribution is found to be lognormal, as is typical of aggregation/condensation sources [18]. Typical average sizes were $\bar{N} \approx 7500$ and 9000 atoms during $^7$Li and $^6$Li measurements, respectively, and the full-width-at-half-maximum was of the same magnitude, $\Delta N \simeq \bar{N}$ . Converting this distribution into a distribution of particle radii gives average values of 3.2 and 3.4 nm, respectively. Here $R = r_s a_0 N^{1/3}$ , $a_0$ is the Bohr radius and $r_s$ = 3.25 is the Wigner-Seitz electron density parameter of lithium.

After a free flight path of approximately 10 ms duration, the nanoparticle beam is photoionized by a Xe or Hg-Xe arc lamp through a monochromator (UV wavelength calibration accurate to <1 nm, as checked at the beginning of every experiment, and bandwidth <0.4 nm). Because the ionization process is a single-photon one, the detected cation count (normalized to both photon and particle fluxes) is equivalent to the photoelectron yield $Y(h\nu)$, where $h\nu$ is the photon energy. It has been found (see [16] and references therein]) that for nanocluster particles the yield curve closely follows the so-called Fowler formula [19-21] describing photoemission from metal surfaces:

$$\log\left(\frac{Y}{T^2}\right) = B + \log f\left(\frac{h\nu - I}{k_B T}\right) \tag{2}$$

Here $I$ is the ionization energy, $B$ is a temperature-independent coefficient incorporating fundamental constants and instrumental parameters, and $f$ is an integral over a portion of the Fermi-Dirac distribution function (counting those electrons which are moving toward the surface with sufficient energy to transition into the continuum after absorbing a photon).

A collection of photoionization yield curves $Y(h\nu)$ obtained by using tunable light are shown in Fig. 1(a) for the full range of investigated nanoparticle temperatures. All profiles can be accurately represented by the universal Fowler function, Eq. (2). Fig. 2 confirms this by showing Fowler fits to all data points for both $^6$Li and $^7$Li, collected at all nanoparticle temperatures. The fits are optimized by randomly sampling points from each dataset 2000 times, see [16] for details. From these fits, the nanoparticle ionization energies $I$ are determined for every given temperature.

Since the nanoparticle beam contains a distribution of sizes, and the average of the distribution shifts slightly between measurement runs, it is useful to represent the data in a consistently scaled way. We do this by extrapolating all ionization energies $I$ to the bulk work function limit $W$ by means of the scaling relation

$$I(T) = W(T) + \alpha \frac{e^2}{R}, \tag{3}$$

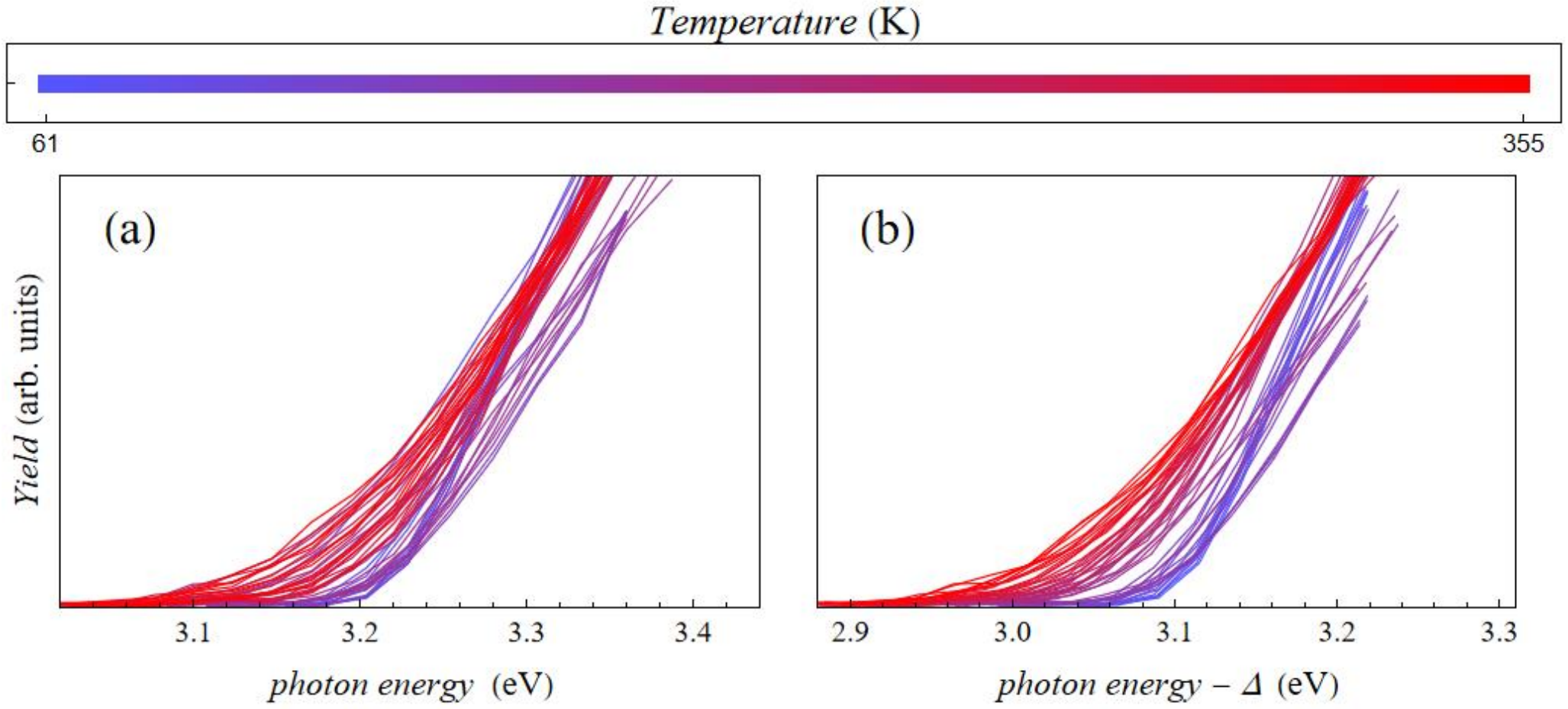

FIG. 1. (a) Raw photoionization yield curves for $^{7}$Li nanoparticles. Each curve is made up of line segments connecting individual data points separated by 3 nm wavelength steps. These data points within each curve are the average of 10 datasets, and for every selected nanoparticle temperature $T$ three or four such (color coded) curves are shown. Hence the plot encodes 30-40 data sets for each temperature setting. (b) When adjusted for particle size variations using Eq. (4), the yield curve positions display consistency across the temperatures. It becomes apparent that as $T$ increases, the tail due to thermally excited photoelectrons extends deeper below the work function threshold.

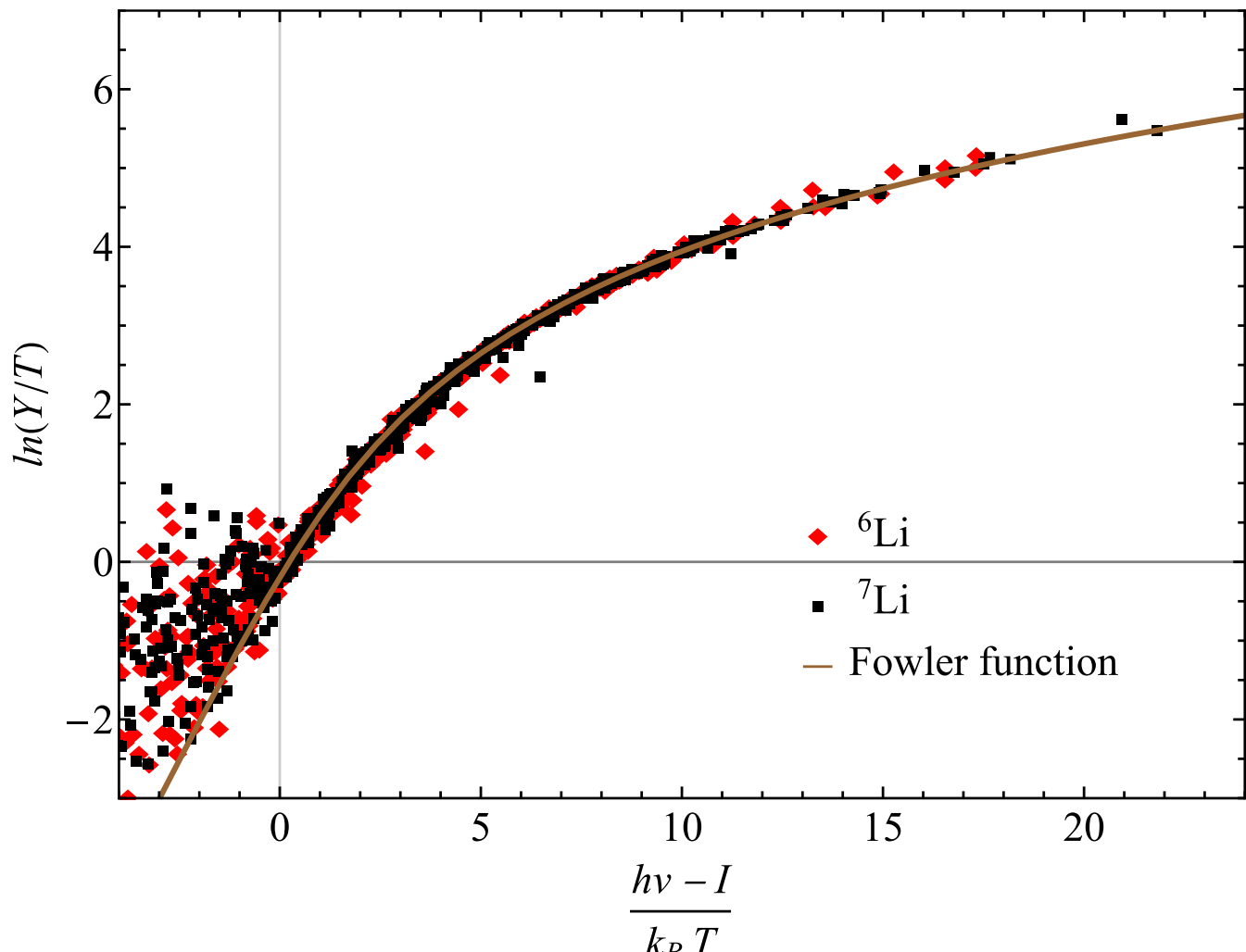

FIG. 2. Fowler plot encompassing the complete $^{6}$Li and $^{7}$Li data sets. For each data set the ionization energy $I$ is fitted so as to make the set's points line up along the function curve. At high temperatures the agreement is very good for points lying both above and below the threshold, the latter signal deriving from thermally excited electrons. For nanoparticles that are cold, the agreement for $h\nu > I$ is also excellent, while the scatter at $h\nu < I$ derives purely from background noise because, as can be seen from Fig. 1, the actual photoelectron signal below threshold becomes very small.

Here $R$ is the particle radius defined above, and the experimental value [22] of the coefficient $\alpha$ for Li is 0.31–0.33 [23].

In practice, the scaling relation is implemented according to a size correction $\Delta$ calculated by convoluting Eq. (3) with the concurrently determined particle size distribution [16,25]:

$$W = I - \Delta = I - \alpha \frac{e^2}{r_s a_0} \frac{\sum_N f(N) N^{-1/3}}{\sum_N f(N)} . \tag{4}$$

Fig. 1(b) shows that when the photoelectron yield curve positions are adjusted for the shift $\Delta$, thereby subtracting out the temperature-induced variations in the intensities $f(N)$, the evolution of these profiles with temperature becomes manifest and consistent.

## III. TEMPERATURE DEPENDENCE OF THE WORK FUNCTION. ISOTOPE EFFECT

Fig. 3 shows the key outcomes of the experiment. First and foremost, the data reveal that the temperature variation of the $^7$Li and $^6$Li work functions are distinct, manifesting a "work function isotope effect." To the best of our knowledge, this is the first report of a dependence of $W$ on the isotopic mass.

Furthermore, the temperature dependence of lithium $W$ is pronouncedly more nonlinear than for other alkali metals (cf. [17,27], and Fig. 6). The reason is that its thermal expansion coefficient (see the inset of Fig. 3 [28]) varies with temperature more rapidly than that of the heavier alkalis. This derives from the fact that the Debye temperatures of the latter are significantly smaller, and therefore their heat capacities – and consequently their thermal expansion coefficients, related via the Grüneisen formula – remain relatively unchanged until much lower temperatures [29].

The inset of Fig. 3 also shows that $^7$Li has a greater thermal expansion coefficient than $^6$Li, which is consistent with the steeper temperature dependence of its work function.

That said, upon closer examination it becomes clear that in lithium metal the temperature variation of $W(T)$ entails dynamics beyond an expansion-related change in the density of the electron gas. The anticipated significance of additional quantum-mechanical effects was already alluded to in the Introduction. The evidence derives from a pronounced mismatch between the Li $W(T)$ data and an electron gas-based image charge model which is successful for free electron-like metals such as Na and K.

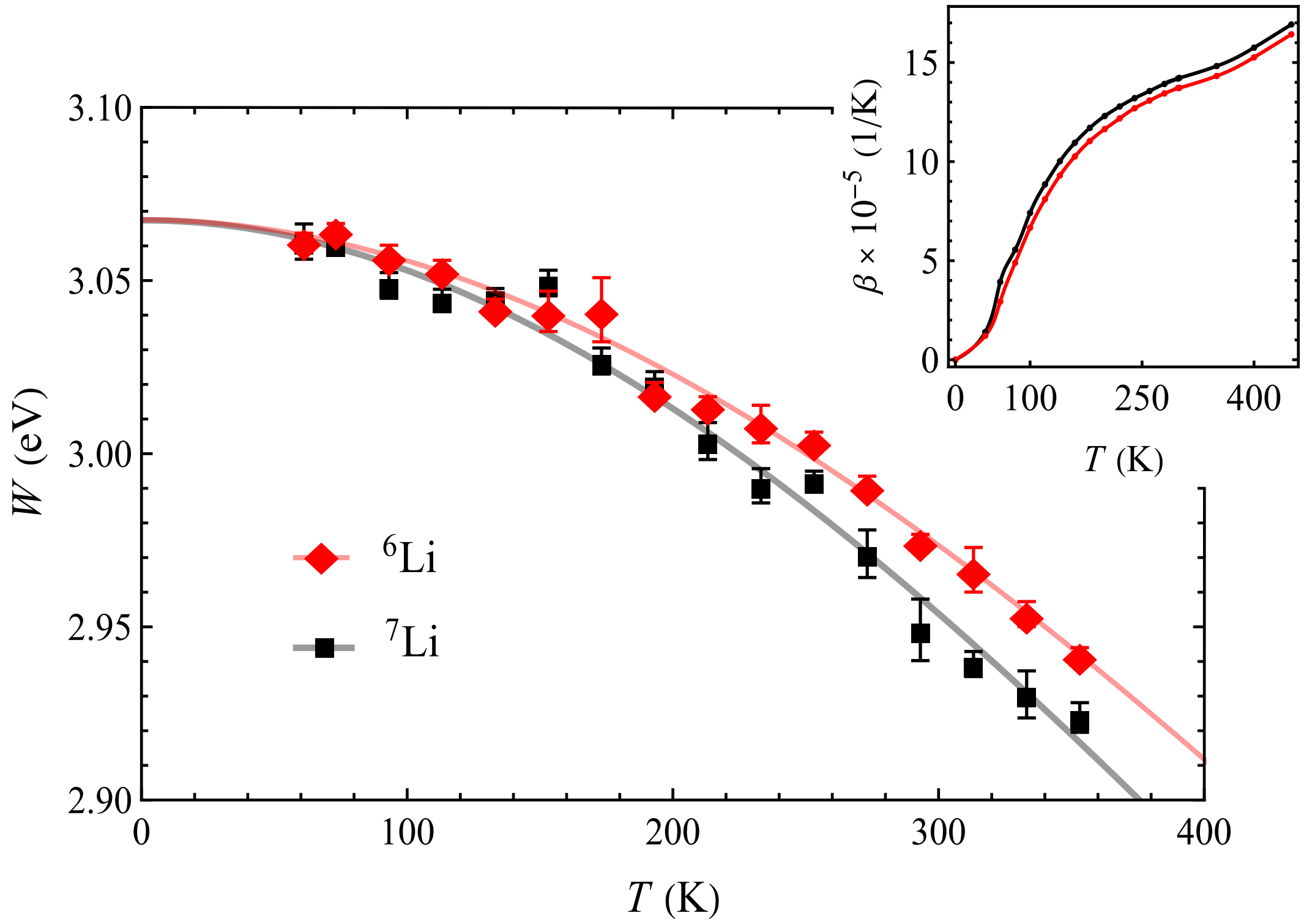

FIG. 3. Experimentally determined work functions of lithium isotopes, $^7$Li (squares) and $^6$Li (diamonds). The error bars represent the standard error, typically ~0.3%, derived from three individual photoionization measurements at each temperature. (Below 150 K, where the variation of $W$ becomes small, five individual measurements at each temperature were performed.) The solid lines are least-square polynomial fits to the data. The inset shows the volume thermal expansion coefficients of $^7$Li (black top line) and $^6$Li (red bottom line) based on data from [28].

In this model the work function is treated as the energy required to remove a charge from a distance $d$ outside the planar surface to infinity, where $d$ is on the order of the screening length in the electron gas and is therefore fully defined by its density (see the discussion and references in [30,31]). Based on this picture and on the tabulated thermal expansion coefficients, the temperature dependence of the work functions of polycrystalline metals can be evaluated, including accounting for exchange and correlation effects [32]. As described in Appendix B, the results are in very good quantitative agreement with the data on other alkalis such as Na and K, signifying that for these metals the electron gas density is indeed the dominant factor.

However, when we repeat the same calculation for Li by using its experimental thermal expansion data (from the inset of Fig. 1) we find a very substantial discrepancy: the experimental $W(T)$ curves are much steeper than expected from this approach. This is clearly seen in Fig. 4. Neither the magnitude of the slopes nor the difference between the isotopes is adequately reproduced by the electron-gas model.

Since the electron mass enters the model calculation as a parameter, we explored whether the discrepancy may be resolved by employing an effective mass $m^*$. As mentioned in the Introduction, lithium's strong ion core potential enhances $m^*$ relative to the free-electron value significantly more than for the other alkali metals. Therefore, the plot in Fig. 4 shows two additional possibilities [33]: band effective mass $m^* \approx 1.3\ m$ [35,36] and thermal effective mass $m^* \approx 2.2m$ [8]. But these also cannot match either the curvature of the experimental plot or the magnitude of the isotope splitting.

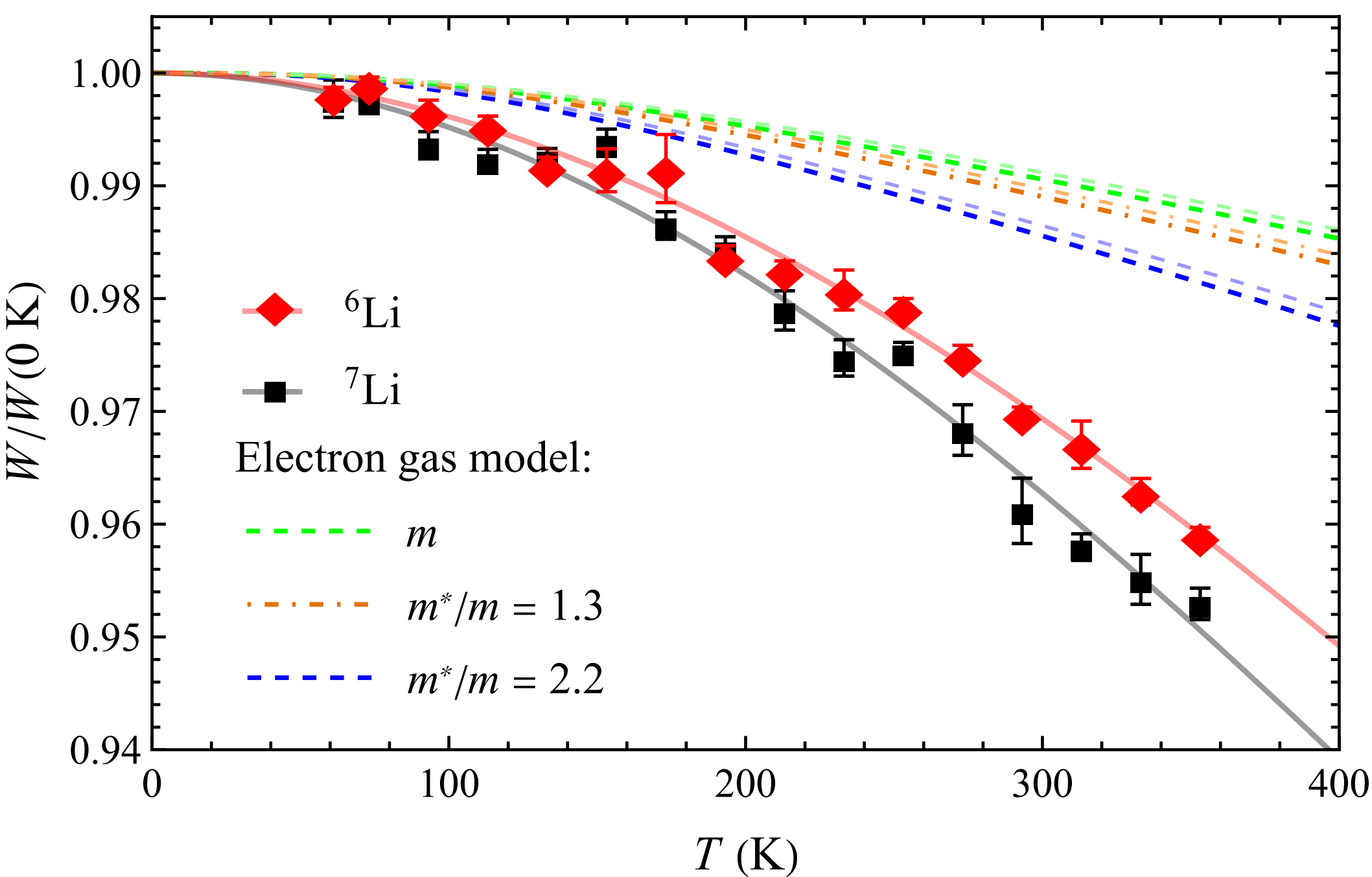

FIG. 4. Fractional shift of the work function with temperature. The data points and solid lines are taken from Fig. 3. The dashed lines follow the formalism of the electron gas model [32], computed using the thermal expansion data from the inset of Fig. 3 (each close-lying pair of lines corresponds to the two lithium isotopes) and three potential options for the electron mass parameter (see text for details). None are able to reproduce the observed temperature dependence of the work function, therefore this dependence derives not only from the change in electron density upon thermal expansion but also from the influence of lattice dynamics.

Hence lattice impact on the electronic work function extends beyond changing the electron gas density via thermal expansion. This raises an important challenge for theory: to quantitatively account for the influence of thermal and zero-point atomic vibrations, including the isotope effect, on the electronic work function of lithium. As pointed out already in the classic review [9], there are at least half a dozen effects which can contribute to the dependence of $W$ on temperature. These include, for example, the fact that the effective electrostatic potential experienced by the conduction electrons is altered by the thermal vibrations of the atoms. In a manner of speaking, this may be viewed as an interesting example of *dynamical* electron-phonon coupling modifying a *static* electronic spectral property (the work function).

A model endeavoring to approximate this effect by replacing the ionic pseudopotential radius by a thermally dilated value [37] did predict a steeper $W(T)$ dependence for Li, indicating that the ionic potential correction is indeed relevant. However, the same calculation strongly overestimated the thermal shift of $W$ for Na and overestimated it for Al [32], for both of which the electron-gas image charge model discussed above was quite successful [17,38. Therefore it is evident that there remains a strong need for a consistent and systematic microscopic treatment.

## IV. THE LOW TEMPERATURE LIMIT OF WORK FUNCTION

At low temperatures, the freezing out of the vibrational degree of freedoms decreases the thermal expansion. This translates into a weaker temperature dependence of the work function at low temperature, as observed in Fig 3. The $W(T \rightarrow 0)$ values extrapolated from the experimental curves are the same for both isotopes: 3.068 eV ± 0.003 eV for $^{6}$Li and 3.068 eV ± 0.004 eV for $^{7}$Li.

The value recommended in the comprehensive literature review [39] for the work function of polycrystalline $^{7}$Li is 2.90 ± 0.03 eV. Viewed as a room-temperature value, this is in close agreement with the data plotted in Fig. 3, validating the accuracy of the present experiment.

The onset of flattening of the $W(T)$ curves aligns well with the temperature at which the thermal expansion shown in the inset experiences a steep decline, reinforcing the correlation between the electronic work function and the vibrational degrees of freedom.

The actual fact that $dW/dT \rightarrow 0$ as $T \rightarrow 0$ can be supported on general grounds, following the argument given in [9]. Starting with Eq. (1), one writes

$$\frac{dW}{dT}=-e\frac{d\left(\Delta\varphi\right)}{dT}-\frac{d\bar{\mu}}{dT} \tag{5}$$

The first term on the right-hand side is defined by electrostatics, and its variation is caused by thermal expansion. Since the thermal expansion coefficient itself scales with the heat capacity of the material, it must vanish for $T \rightarrow 0$ (this can be seen in the inset in Fig. 3). It follows that $d(\Delta\phi/dT) \rightarrow 0$. As regards the second term, one uses the Maxwell relation

$$\left(\frac{\partial\bar{\mu}}{\partial T}\right)_{P,N}=-\left(\frac{\partial S}{\partial N}\right)_{T,P} \tag{6}$$

For solids entropy vanishes as $T \rightarrow 0$, hence $\partial S/\partial N$ tends to zero and so does the derivative of the chemical potential. Hence in total $dW/dT \rightarrow 0$.

The behavior of the work function in Fig. 3 conforms to this behavior. This establishes that there is no extensive residual entropy contained within the nucleated nanoparticles and supplies an elegant illustration of the Third Law of Thermodynamics.

## V. CONCLUSIONS

Accurate measurements of the work function of lithium metal, enabled by the use of high purity nanoparticles in a beam, reveal that its temperature dependence exhibits an isotope effect. The detection of this difference between $^{7}$Li and $^{6}$Li demonstrates that the work function can serve as a probe of the interplay between the electronic and structural degrees of freedom.

The temperature variation of the lithium work function is significantly steeper than predicted by a model that incorporates only thermal expansion of the electron gas, which is sufficient for describing the behavior of metals such as Na and K. This implies that the effect of lattice dynamics on the work function of lithium metal joins the set of other properties of this material that are quantum-mechanical in nature, and is in need of a comprehensive microscopic understanding.

The approach illustrated here can be extended to lower temperatures and to associated systems. For example, bulk $^{7}$Li undergoes a martensitic transition below approximately 77 K, while the behavior of $^{6}$Li is quite different [3]. An isotope-resolved work function measurement taken at small temperature steps and extending to lower nanoparticle temperatures may yield evidence of analogous phenomena at the nanoscale. Another interesting possibility would be to use a dual-oven condensation source to produce beams of metal-fullerene nanoparticles $M_x(C_{60})_n$ [40,41], using accurate photoionization measurements to identify the amount of charge transfer and the formation of packing structures in these prototype fulleride compounds.

## ACKNOWLEDGMENTS

We would like to thank Prof. Shanti Deemyad for a useful discussion. This research was supported by the U.S. National Science Foundation under Grant No. DMR-2003469.

## APPENDIX A: SIZE DISTRIBUTION OF THE GAS-PHASE NANOPARTICLES

Mass spectra are recorded immediately before and after each photoelectron threshold data set measurement. They are obtained by one-photon above-threshold ionization of the nanoparticles, using 355 nm third-harmonic pulses from a Nd-YAG laser focused into the extraction region of a linear Wiley-McLaren time-of-flight (TOF) mass spectrometer [16]. As mentioned in Sec. II, the size distribution is well described by a lognormal function. Fig. 5 shows a representative mass spectrum *f*(*N*).

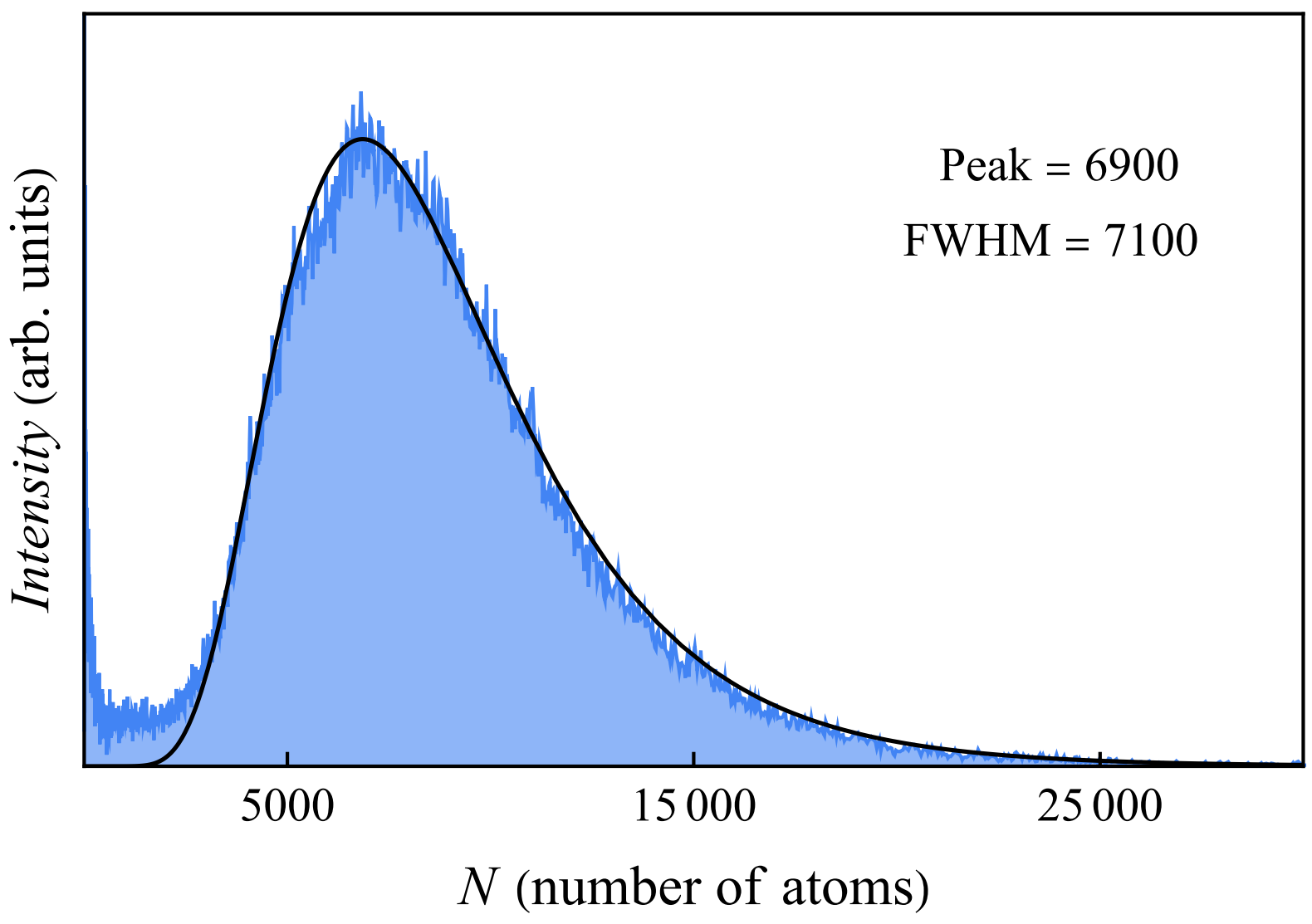

FIG. 5. Size distribution of $^7$Li nanoparticles thermalized to 80$^\circ$ C. The data are well described by a lognormal distribution (black envelope). The wing at the low end of the distribution is an artifact caused by TOF laser-induced fragmentation [16] and is not included in the calculation in Eq. (4).

## APPENDIX B: ELECTRON GAS MODEL

The temperature dependence of the electronic contribution to the alkali bulk metal work function is modeled in [32] as an effect that derives solely from thermal expansion, using the following expression derived in [30]:

$$W(T) = \frac{37.38}{\left[r_s(T)\right]^{3/2}\left[E_F(T)\right]^{1/2}}, \tag{B1}$$

where the Fermi energy $E_F$ is expressed in eV.

We calculated the temperature dependence of the Wigner-Seitz radius $r_s(T)$ numerically, making use of the tabulated volume thermal expansion coefficients $\beta(T)$ of K [42], Na [43], and $^6$Li and $^7$Li [28].

The authors of Ref. [32] propose that the best evaluation of the temperature variation of $E_F$ in the denominator of Eq. (S.2) is obtained by augmenting it with exchange and correlation effects. Therefore the model employs $dE_F/dT = d\mu/dT$, where the electron gas chemical potential, in eV, is

$$\mu(r_s) = \frac{50.0}{r_s^2} - \frac{16.6}{r_s} - \frac{12.0}{r_s + 7.8}\left(1 + \frac{r_s}{3(r_s + 7.8)}\right). \tag{B2}$$

The temperature variation of this expression can be obtained from $r_s(T)$ discussed above.

The two parameters are then combined in the denominator of Eq. (S.2), yielding the model theoretical curve for the thermal shift of the work function. Ref. [32] provided a listing of parabolic fits to $W(T)$, but since $\beta(T)$ is highly nonlinear when followed over a wide temperature range (cf. the inset in Fig. 3), we plot here the full numerically integrated functions.

Note also that the first term in Eq. (B2) represents the numerical value of the free electron gas Fermi energy, and is therefore inversely proportional to the electron mass. By incorporating an effective mass factor into this term, the scaling of $dW/dT$ with $m^*$ can be explored, as discussed in Sec. III.

Fig. 6 affirms that the present electron model provides excellent agreement with our previous data, extracted from analogous experiments with Na and K nanoparticles of similar sizes [17]. This is consistent with their characterization as nearly free-electron-gas metals, and the model captures the evolution of their work functions as being primarily due to the thermal expansion of the solid.

However, as demonstrated in Fig. 4, this proves to be insufficient for explaining the behavior of lithium metal: neither its work function's temperature dependence nor its strong isotope effect.

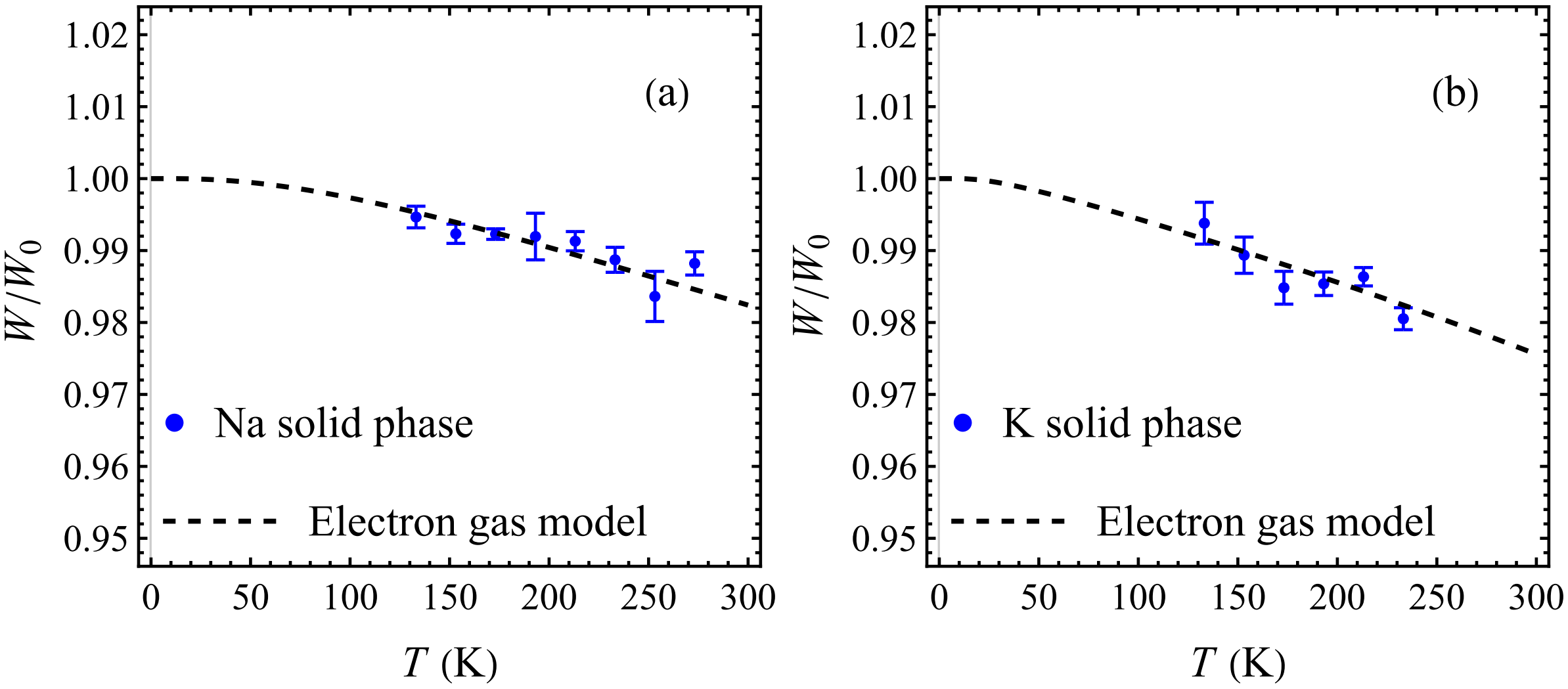

FIG. 6. Temperature variation of the metal work function predicted by the electron gas model (dashed line) compared to the experimental data for Na and K at temperatures below the nanoparticles' melting points [17]. $W(T)$ is plotted relative to its value at 0 K, $W_0$. Fitting the model electron gas curve to the data yields $W_0$ = 2.795 eV for Na and $W_0$ = 2.375 eV for K, which is 0.3% and 0.6%, respectively, below the straight-line extrapolation of the data points. As mentioned in Sec. III, the $W(T)$ plots for these metals retain their nearly straight-line shapes to much lower temperatures than Li (cf. main text figures) because the Debye temperature of the latter (300–400 K) is substantially higher than that of Na (≈160 K) and K (≈90 K) [8].